\documentclass[twocolumn]{aastex62}

\newcommand{\planck}{\textsl{Planck}}
\newcommand{\Planck}{\textsl{Planck}}
\newcommand{\lcdm}{\ensuremath{\Lambda\mathrm{CDM}}}
\newcommand{\wmap}{\textsl{WMAP}}
\newcommand{\WMAP}{\textsl{WMAP}}

\graphicspath{{./}{figures/}}

\shorttitle{Quantifying the Degeneracy Between the Matter Density and Hubble Constant}
\shortauthors{Kable et al.}

\begin{document}

\title{Quantifying the CMB Degeneracy Between the Matter Density and Hubble Constant in Current Experiments}

\correspondingauthor{Joshua A. Kable}
\email{jkable2@jhu.edu}

\author[0000-0002-0786-7307]{Joshua A. Kable}
\affil{Johns Hopkins University \\
3400 North Charles Street \\
Baltimore, MD 21218, USA}

\author{Graeme E. Addison}
\affiliation{Johns Hopkins University \\
3400 North Charles Street \\
Baltimore, MD 21218, USA}

\author{Charles L. Bennett}
\affiliation{Johns Hopkins University \\
3400 North Charles Street \\
Baltimore, MD 21218, USA}



\begin{abstract}

We revisit the degeneracy between the Hubble constant, $H_0$, and matter density, $\Omega_m$, for current cosmic microwave background (CMB) observations within the standard \lcdm\ model. We show that \planck, Wilkinson Microwave Anisotropy Probe (\wmap), South Pole Telescope (SPT), and Atacama Cosmology Telescope Polarimeter (ACTPol) temperature power spectra produce different values of the exponent $x$ from minimizing the variance of the product $\Omega_mH_0^x$. The distribution of $x$ from the different data sets does not follow the Markov Chain Monte Carlo (MCMC) best-fit values for $H_0$ or $\Omega_m$. Particularly striking is the difference between \planck\ multipoles $\ell\leq800$ ($x=2.81$), and \wmap\ ($x = 2.94$), despite very similar best-fit cosmologies. We use a Fisher matrix analysis to show that, in fact, this range in exponent values is exactly as expected in \lcdm\ given the multipole coverage and power spectrum uncertainties for each experiment.  We show that the difference in $x$ from the \planck\ $\ell \leq 800$ and \wmap\ data is explained by a turning point in the relationship between $x$ and the maximum effective multipole, at around $\ell=700$. The value of $x$ is determined by several physical effects, and we highlight the significant impact of gravitational lensing for the high-multipole measurements. Despite the spread of $H_0$ values from different CMB experiments, the experiments are consistent with their sampling of the $\Omega_m-H_0$ degeneracy and do not show evidence for the need for new physics or for the presence of significant underestimated systematics according to these tests. The Fisher calculations can be used to predict the $\Omega_m-H_0$ degeneracy of future experiments.

\end{abstract}



\section{Introduction} \label{sec:intro}
The \lcdm\ model has been the standard model of cosmology for  nearly two decades and successfully explains a range of observations, including abundance of primordial elements produced by Big Bang Nucleosynthesis \citep[see, e.g., review by][]{cyburt/etal:2016}, baryon acoustic oscillations in the CMB and clustering of large-scale structure \citep[e.g.,][]{alam/etal:2017}, and current accelerating expansion probed by Type Ia supernovae \citep[see][for the latest results]{scolnic/etal:2018}. In recent years, cosmological experiments have become increasingly precise and tensions have emerged within \lcdm. The most pronounced tension is the disagreement over the value of the Hubble constant, $H_0$ \citep[e.g.,][]{bernal/etal:2016,planck/6:2018,riess/etal:2018,addison/etal:2018,Lemos/etal:2018}.

The fundamental question today is whether new physics is needed, or the precision of some data is over-estimated. This motivates us to take a closer look at the role of parameter degeneracies in $H_0$ determinations from CMB temperature anisotropy theory and observation. The CMB forms the cornerstone of modern cosmology because it precisely constrains all \lcdm\ parameters simultaneously. Using the CMB anisotropy and assuming a cosmological model, $H_0$ can be indirectly calculated.  However, it has long been known that there is a degeneracy between $H_0$ and $\Omega_m$ when fitting to CMB data \citep{Zaldarriaga/etal:1997}. The $\Omega_m-H_0$ degeneracy can be expressed in terms of the spacing of the acoustic peaks  $\theta_*$ \citep{Hu/etal:2001,percival/etal:2002}. Percival et al. showed that for temperature data in flat universe models the acoustic horizon angle is 
\begin{equation}
\theta_* = \frac{r_*}{D_*},
\end{equation}
where $r_*$ is the comoving sound horizon at the surface of last scattering, approximately given by 
\begin{equation}
r_* = \frac{1}{H_0\sqrt\Omega_m}\int_0^{a_*} \frac{c_s da}{\sqrt{a+a_{eq}}},
\end{equation}
and $D_*$ is the comoving angular diameter distance to the surface of last scattering, approximately given by
\begin{equation}
D_* = c \int_0^{z_*} \frac{dz}{H_0\sqrt{\Omega_m (1+z)^3+1-\Omega_m}} \propto \frac{\Omega_m^{-0.4}}{H_0}
\end{equation}
 for a flat universe where $\Omega_{\Lambda} = 1 - \Omega_m$. Here `$*$' refers to recombination, $a_{eq}$ is the scale factor of matter-radiation equality, and $c_s$ is the speed of sound given by $c_s = c / (3(1+\frac{3 \rho_b}{4\rho_{\gamma}}))^{0.5}$ where $\rho_b$ and $\rho_{\gamma}$ are the baryon and photon densities respectively. Assuming that $c_s$ is constant over the interval $(0,z_*)$, then
\begin{equation}
\theta_* \propto \frac{\sqrt{1+a_{eq}z_*}-\sqrt{a_{eq}z_*}}{\Omega_m^{0.1}\sqrt{1+z_{*}}}.
\end{equation}
Near a fiducial value of $\Omega_mh^2 = 0.147$, which sets the redshift of matter-radiation equality, and assuming that the redshift of the surface of last scattering is invariant to \lcdm\ parameters, Percival et al. derived
\begin{equation}
\frac{\partial \log(\theta_*) }{\partial  \log(\Omega_m)} \bigg |_{h} = 0.14
\end{equation}
\begin{equation}
\frac{\partial  \log(\theta_*) }{\partial  \log(h)} \bigg |_{\Omega_m} = 0.48
\end{equation}
 meaning $\theta_* \propto \Omega_mh^{3.4}$. This describes the expected degeneracy for a CMB measurement of fixed peak spacing. 

In addition to the peak spacing, CMB data also provide a constraint on the peak heights, which depend on the physical matter density, $\Omega_mh^2$. Percival et al. conclude that the combination of $\Omega_mh^{3.4}$ from the peak spacing and $\Omega_mh^2$ from the peak heights gives an approximate degeneracy of $\Omega_mh^{\sim 3}$. 

While there is a degeneracy between $\Omega_m$ and $H_0$, the power law fit is only an approximation. Moreover, the degeneracy itself is sensitive to changes in the other parameters. In particular, changing $\Omega_mh^2$ changes the heights of acoustic peaks, which shifts the values of the physical baryon density, $\Omega_bh^2$ and scalar tilt, $n_s$, to compensate. Because of this, different fiducial values of $\Omega_bh^2$  and $n_s$ can change the $\Omega_m-H_0$ degeneracy. Independent of the other parameters, the degeneracy is partially broken by physical effects such as the late-time integrated Sachs-Wolfe (ISW) effect and gravitational lensing \citep{howlett/etal:2012}. Further, unexpected new physics that is related to $\Omega_m$ or $H_0$ can easily affect this relation. 
\begin{figure}[!tbp]
  \centering
  \begin{minipage}[b]{0.4\textwidth}
  \hspace*{-1cm}
 \includegraphics[width=3.464in]{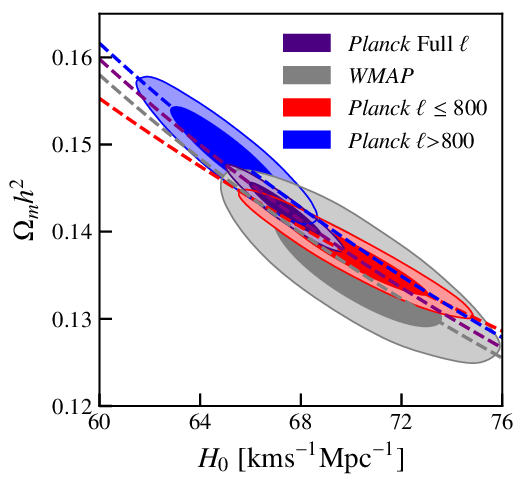}
  \end{minipage}
  \hfill
  \begin{minipage}[b]{0.4\textwidth}
    \hspace*{-1cm}
    \includegraphics[width=3.464in]{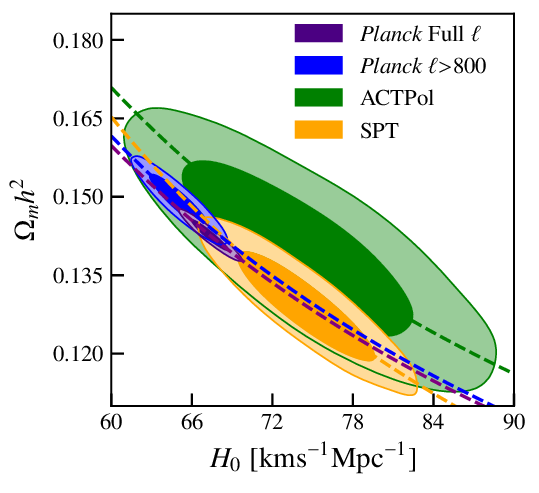}
  \end{minipage}
  \caption{The $\Omega_mh^2-H_0$ degeneracy is easily seen for CMB experiments  incorporating only temperature data and a prior of $\tau = 0.07\pm 0.02$.  In addition to the full \planck\ data set, we include the \planck\ data set split into two parts at $\ell = 800$. We have split the Figure into two panels to illustrate CMB data sets that measure large angular scales in the top panel and small angular scales in the bottom panel. In the top panel, we include \planck\ $\ell>800$ to show the tension between the two halves of the \planck\ data.  The CMB data sets prefer different degeneracy directions shown as dashed lines (see Section 2)  as visible by eye for \wmap\ and \planck\ $\ell\leq800$. In this paper, we investigate whether degeneracy directions are consistent with \lcdm\ expectations.  
\label{fig:f1}}
\end{figure}

The CMB $\Omega_m-H_0$ degeneracy is shown in Figure~1. In this plot, 1 and $2\sigma$ confidence contours are taken from Markov Chain Monte Carlo (MCMC) fits to temperature (`TT') power spectra from \planck\ 2015\footnote{The \planck\ 2015 and 2018 TT power spectra were shown to be in good agreement in Section 3.6 of \cite{planck/6:2018}.} \citep{planck/13:2015}, \WMAP\ \citep{hinshaw/etal:2013}, SPT \citep{story/etal:2013}, and ACTPol \citep{louis/etal:2017}. These chains were obtained using the \texttt{CosmoMC}\footnote{\url{https://cosmologist.info/cosmomc/}} package \citep{lewis/bridle:2002} and public foreground-marginalized likelihood codes\footnote{Available from the \planck\ Legacy Archive \url{http://pla.esac.esa.int/pla/\#home} and Legacy Archive for Microwave Background Data Analysis (LAMBDA) \url{https://lambda.gsfc.nasa.gov/product/}}. Additionally, we show results from the \Planck\ TT spectrum split at $\ell = 800$. This splits the \planck\ constraining power roughly in half \citep{addison/etal:2016,planck/51:2017} and allows for comparisons of \planck\ $\ell\leq800$ with \wmap\ and \Planck\ $\ell>800$ with small-scale measurements from ACTPol and SPT. A common prior on the optical depth of $\tau=0.07\pm0.02$ was applied in all cases \citep{weiland/etal:2018}.

The mean and 68$\%$ confidence intervals for  $H_0$ and $\Omega_mh^2$ from the MCMC chains used in Figure~1 are shown in Table~1. Using the  \planck\ Full $\ell$ range produces a value of $H_0$ between the lower value from \Planck\ $\ell>800$ and the higher value from \Planck\ $\ell\leq800$. Note that, despite both examining high-$\ell$ modes, \Planck\ $\ell>800$ and SPT have a 2.6$\sigma$ $H_0$ tension. While \Planck\ $\ell\leq800$ and \WMAP\ predict very similar values for $H_0$, the exact degeneracy direction (orientation of the contours) is different. We restrict analysis in this work to TT spectra to compare with Percival et al., but the degeneracy is still present in polarization data. The large uncertainty for ACTPol is a result of not including polarization data. 

We provide additional tests and insights into the $\Omega_m-H_0$ degeneracy by asking whether the degeneracy directions are consistent with expectations and whether any experiments are outliers. In Section~2, we outline our method for fitting the power law to the MCMC points. In Section~3, we show the power law fits from the MCMC and employ a Fisher analysis to calculate the expected range of power law exponents for each experiment. In Section~4, we extend the Fisher analysis to understand the power law fits observed for different \planck\ multipole cuts and \WMAP, as well as how lensing affects the power law fit. In Section~5, we provide conclusions. 
\begin{deluxetable}{ccc}
\tabletypesize{\footnotesize}
\tablewidth{0pt}

 \tablecaption{Mean values and 68$\%$ confidence intervals for $H_0$ and $\Omega_mh^2$ for the MCMC chains used in Figure~1, which includes a prior of  $\tau=0.07 \pm 0.02$ as described in the text. \label{tab:best-fit}}

 \tablehead{
 \colhead{Experiment} & \colhead{$H_0$  [km $\textrm{s}^{-1}\textrm{Mpc}^{-1}$]} & \colhead{$\Omega_mh^2$}
 }

 \startdata 
 \Planck\ Full $\ell$  & 67.4 $\pm$ 1.0 & 0.143 $\pm$ 0.002   \\
 \WMAP  & 70.2 $\pm$ 2.2 & 0.136 $\pm$ 0.005 \\
 \planck\ $\ell\leq800$  & 70.1 $\pm$ 1.9 & 0.137 $\pm$ 0.003 \\
 \planck\ $\ell>800$  & 65.0 $\pm$ 1.5 &  0.149 $\pm$ 0.003 \\
 ACTPol  & 74.3 $\pm$ 5.7 & 0.140 $\pm$ 0.011 \\
 SPT  & 74.5 $\pm$ 3.4 & 0.129 $\pm$ 0.007  \\
 \enddata
 \tablenotetext{}{The distance ladder measures $H_0$ = 73.5 $\pm$ 1.7 km $\textrm{s}^{-1}\textrm{Mpc}^{-1}$ \citep{riess/etal:2018}.}

 \vspace{-0.5cm}

\end{deluxetable}

\section{Calculating the power law exponent}

We perform calculations for the data sets shown in Figure~1 (and Table~1): \Planck\ Full $\ell$, \Planck\ $\ell\leq800$, \Planck\ $\ell>800$, \WMAP, ACTPol, and SPT. Again, we restrict the MCMC to TT power spectra to compare with Percival et al. To facilitate comparison with Fisher matrix predictions we only used multipoles where the likelihood for the power spectrum $C_{\ell}$'s is approximately Gaussian, and for this reason we removed $2\leq\ell<30$ from the \planck\ Full $\ell$, \planck\ $\ell\leq800$, and \wmap\ analyses. For \Planck\ this has only a minor effect on the power law exponent; however, there is a non-negligible effect on \WMAP, which is discussed in Section 3.  For each experiment, the value of the optical depth was fixed to $\tau = 0.07$. The TT power spectrum with $\ell \geq 30$ is predominately sensitive to the degenerate combination $A_se^{-2\tau}$, so fixing $\tau$ has a small effect on $\Omega_mh^2$ and $H_0$ determinations. We tested the impact of using the \cite{planck/6:2018} value of $\tau = 0.054$ and found that it had a negligible effect on the preferred power law direction compared to the spread from different experiments. The best-fit cosmological parameters resulting from the MCMC runs for each data set are shown in Table~2. 

To calculate the power law fit, we extract the chains of ordered pairs, $(\Omega_m,H_0)$, from the output of the MCMC. This automatically marginalizes over the other parameters. To calculate the relative exponent, we make a linear fit to the log-log chains. We subtracted off the mean values of the log-log chains to reduce the fit to be one dimensional, i.e., the slope of the line. We find the slope that minimizes the variance of the quantity $\log(\Omega_mh^x)$. In particular, if we define $(M,H) = (\log(\Omega_m/\bar{\Omega}_m),\log(H_0/\bar{H}_0))$, then we can define $x$ to be the slope of the direction that maximizes the variance along its axis. Each chain point can be represented in the principal component basis as 
\begin{equation}
P_1 = \frac{1}{\sqrt{x^2+1}}(H - Mx) 
\end{equation}

\begin{equation}
P_2 = \frac{1}{\sqrt{x^2+1}}(Hx + M).
\end{equation}

We find $x$ that minimizes the variance of the $P_1$ chain points, which is the root of the quadratic equation

\begin{equation}
-\sum_i [H_iM_i x^2 - (H_i^2 - M_i^2) x -  H_iM_i ]= 0,
\end{equation}
where the sums are over all of the chain points. The two roots correspond to a maximum and a minimum, and we take the minimum. 

\section{Results and Fisher calculations}

The preferred power law exponents from the $\Omega_mh^2$ and $H_0$ chains for each data set are shown as black points in Figure~2. Values range from 2.81 for \planck\ $\ell\leq800$ to 3.17 for SPT. It is unclear, using only MCMC results, whether these differences are significant. By subdividing the chains we found that the uncertainty associated with the numerical calculation of the power law exponent from a finite number of chain points is negligible compared to the spread across experiments. The uncertainty associated with the cosmic variance and noise for a particular experiment cannot be easily obtained from a single MCMC run, however.

Interestingly, the MCMC differences between experiments do not follow a trend set by the best-fit cosmology's value for $H_0$. SPT and \planck\ $\ell > 800$ both prefer power law exponents greater than 3 despite estimating very different values for $H_0$. Meanwhile \planck\ Full $\ell$, \planck\ $\ell \leq 800$, and \wmap\ estimate similar $H_0$ values as seen in Table~2, yet prefer power law exponents that are different from one another and are each smaller than for \planck\ $\ell > 800$. 

The difference in the degeneracy directions between \planck\ $\ell\leq800$ and \wmap\ is particularly surprising at first considering \cite{huang/etal:prep} found their power spectra to agree within $1\sigma$. Further, there is good agreement between cosmologies as seen in Figure~1 and Table~1. Differences in other parameters that $\Omega_mh^2$ and $H_0$ are correlated with, such as $\Omega_bh^2$ or $n_s$, cannot explain this discrepancy because both data sets favor similar values. There are no obviously different physical effects in play. For example, neither \planck\ $\ell\leq800$ nor \wmap\ has much sensitivity to gravitational lensing, and $\ell<30$ is removed in both cases, removing most dependence on the ISW effect \citep{howlett/etal:2012}. Including $\ell < 30$ in the MCMC fit increases the \wmap\ preferred power law exponent to 2.97 while there is only a minimal change to \planck\ $\ell\leq 800$. 

\begin{figure}[!tbp]
  \begin{minipage}[b]{0.4\textwidth}
    \hspace*{-0.4cm}
    \includegraphics[width=3.75in, height=3in]{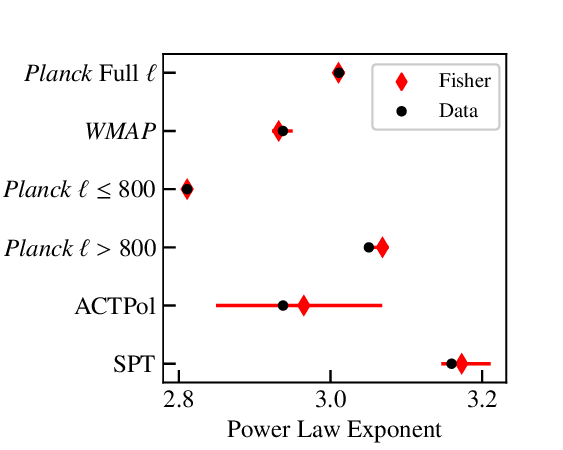}
  \end{minipage}
  \caption{The power law exponents preferred by each data set. The power law exponent $x$ is defined to minimize the variance of $\log(\Omega_mh^x)$. This was calculated for MCMC fits to measured data and for Fisher estimates using the MCMC best-fit cosmologies as fiducial models. Additionally, we used a Fisher analysis to calculate $x$ for 100 cosmologies randomly selected from that experiment's MCMC chain. We then plotted the range corresponding to the middle 68 power law exponents as red lines.  The Fisher analysis seems to reliably replicate the MCMC. The different data sets prefer significantly different power law exponents, but Fisher calculations show that this is expected given the range of multipoles and noise levels across the experiments.  
 \label{fig:general}}
  \end{figure}

Understanding the expected exponents for each data set within \lcdm\ is clearly important. While the different collaborations have shown that tests of \lcdm\ consistency based on, for example, $\chi^2$, are broadly passed, it seems plausible that the value of the $\Omega_m-H_0$ power-law exponent could be more sensitive to some particular type of systematic error, problem in the analysis, or some unexpected new physics.
  
To estimate the expected power-law exponent for each data set, we use a Fisher analysis. This approach has been widely used for forecasting results of cosmological experiments because it is a computationally inexpensive way of calculating parameter uncertainties \citep[e.g.,][]{Heavens:2009,Verde:2010}. This is achieved by assuming that the measured data ($C_{\ell}$'s, or binned values $C_b$) respond linearly to small deviations in the underlying parameters $\theta$. For a Gaussian likelihood 
where the bandpower covariance matrix, $\mathbb{C}_{bb'}$, may be well-approximated as independent of cosmological parameters, the Fisher matrix is given by
\begin{equation}
F_{ij} = \sum_{bb'} \frac{\partial C_b^{th}}{\partial \theta_i} \mathbb{C}^{-1}_{bb'}\frac{\partial C_{b'}^{th}}{\partial \theta_j},
\end{equation}
where $C_{\ell}^{th}$ refers to the theory power spectrum for a given fiducial model. In each case, we use the data covariance matrix provided by each collaboration with foreground parameters marginalized over \citep{dunkley/etal:2013}. The binned power spectrum for each experiment is obtained using binning matrices, $B_{b\ell}$, provided in the public likelihood codes:
\begin{equation}
C_b=\sum_{\ell}B_{b\ell}C_{\ell}.
\end{equation}
Note that the binning matrices from SPT operate on $\ell(\ell+1)C_{\ell}$ rather than $C_{\ell}$.

The derivatives of the $C_l's$ were calculated numerically with a two-sided derivative about a fiducial model and with step sizes of $1\%$ changes in each parameter. For fiducial value $\hat{\theta}_i$ and fractional change $\Delta$,
\begin{equation}
\frac{\partial C_{\ell}^{th}}{\partial \theta_i} = \frac{C_{\ell}^{th}((1+\Delta)\hat{\theta}_i)-C_{\ell}^{th}((1-\Delta)\hat{\theta}_i)}{2\Delta\hat{\theta}_i}.
\end{equation}

While taking the derivative with respect to the $i^{th}$ parameter, we hold fixed the other parameters. We tested the numerical stability of the Fisher matrix by varying $\Delta$ between 0.2$\%$ and $2\%$ and found that changes to the power law exponent were smaller than 1$\%$. 
The inverse of the Fisher matrix is the parameter covariance matrix. We compared the estimated parameter covariance matrix from the Fisher analysis with the parameter covariance matrix formed by the output of the MCMC chains, finding agreement to within $10-15\%$ with most diagonal elements showing $1-5\%$ agreement. While there are differences between the inverse of the Fisher matrix and the MCMC parameter covariance matrix for ACTPol on the order of $15-25\%$, we attribute this to non-Gaussian features  in the data that can be seen in the non-elliptical features present in the ACTPol contours of Figure~1 as well as weaker constraints from our use of only temperature data.  

\begin{deluxetable}{cccccc}[h]
\tabletypesize{\footnotesize}
\tablewidth{0pt}

 \tablecaption{Best-fit \lcdm\ parameters for data sets used for determining power-law exponents in this work, excluding $\ell<30$ and fixing $\tau=0.07$ as described in the text. $H_0$ is in units of km$\textrm{s}^{-1}\textrm{Mpc}^{-1}$. \label{tab:best-fit}}

 \tablehead{
 \colhead{Experiment} & \colhead{$H_0$} & \colhead{$\Omega_bh^2$} & \colhead{$\Omega_ch^2$}& \colhead{$10^9A_s$} & \colhead{$n_s$}
 }

 \startdata 
 \Planck\ Full $\ell$  & 67.07 & 0.0222 & 0.1204 & 2.167 & 0.9630 \\
 \WMAP  & 68.33 & 0.0223 & 0.1163 & 2.137 & 0.9644\\
 \planck\ $\ell\leq800$  & 67.71& 0.0221 & 0.1187 & 2.150 & 0.9588\\
 \planck\ $\ell>800$  & 64.06 &  0.0222 & 0.1291 & 2.213 & 0.9506\\
 ACTPol  & 73.30& 0.0243 & 0.1145 & 2.209 & 0.9474 \\
 SPT  & 74.28 & 0.0231 & 0.1059 & 2.195 & 0.9309\\
 \enddata
 
\end{deluxetable}

While it is possible to analytically calculate the power law fit directly from the Fisher matrix, we instead draw chain points from the inverse of the Fisher matrix to as closely as possible replicate an MCMC run. We used the best-fit cosmologies from the MCMC runs as the fiducial models for the Fisher analysis, and we used \texttt{CAMB}\footnote{https://camb.info/} to take these fiducial models and generate temperature power spectra $C_{\ell}$ \citep{lewis/etal:2000}. 

To quantify the spread in possible preferred power law exponents associated with using different fiducial models in the Fisher analysis, we recomputed the Fisher matrix for a range of cosmologies for each experiment. We chose the range of cosmologies for each experiment by randomly selecting 100 different cosmologies from that experiment's MCMC chains. In general we found that, using a fiducial model with $H_0$ smaller (larger) than the $H_0$ from the MCMC best-fit cosmology results in a smaller (larger) preferred power law exponent. We put these power law exponents in ascending order and plotted the range corresponding to the central 68 values in Figure~2 as a red line. 

The different choices of a fiducial model correspond to the uncertainty associated with the cosmic variance and noise for a particular experiment, meaning that the 68 data points give the Fisher prediction for the possible spread in the values of preferred power law exponent. For all data sets, the MCMC fit to data lies within the central 95 Fisher power law calculations.

Using the power law fits to the MCMC chains, we show that the experiments prefer different power law exponents. Including the Fisher analysis with the uncertainty associated with choice of fiducial model, we illustrate that these differences are significant; however, the preferred power law exponents for each experiment are expected within \lcdm. \textit{This means that given \lcdm\ cosmology, experiments such as \planck\ $\ell \leq 800$ and \wmap\ are not expected to see the same degeneracy direction, implying that the experiments are constraining $\Omega_mh^2$ and $H_0$ differently. This is caused by the different experimental weightings of multipole moments that are captured in the multipole covariance matrix.}

\section{Testing effect of \planck\ multipole range, \wmap\ noise, and lensing}

To examine the shifts in power-law exponent in more detail, we studied the effect of changing maximum multipole moment for both the \planck\ data and Fisher calculations. The red points in Figure~3 were obtained by repeating the \planck\ Fisher analysis but truncating the  $C_{\ell}^{\rm th}$ at maximum multipoles between 550 and 2500 in intervals of $\Delta\ell=10$. The \planck\ $\ell\leq800$ best-fit cosmology was used to compute the derivatives. Figure~3 shows that the power law exponent values ranges from approximately 2.7 to 4.1. Increasing the multipole cutoff from 550 to 700 results in a sharp drop in the power law exponent from around 4 to its minimum of 2.75. This range of multipoles includes the maximum and roll-over of the second acoustic peak and coincides with significant changes in the parameter degeneracies including the degeneracies of $\Omega_mh^2$ and $H_0$ with $\Omega_bh^2$ and $n_s$. 

\begin{figure}[!tbp]
\centering
  \begin{minipage}[b]{0.4\textwidth}
  \hspace*{-1cm}
    \includegraphics[width=3.75in, height=3in]{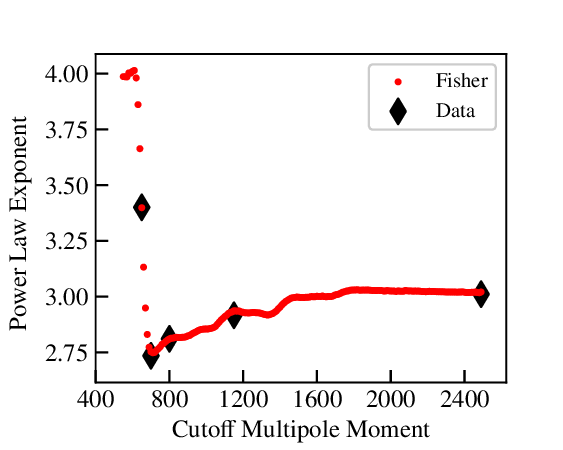}
  \end{minipage}
  \caption{The power law fit exponent for a \Planck-like\ experiment with varying upper multipole cutoff. We fit the exponent $x$ that minimizes the variance of the combination $\log(\Omega_mh^x)$ for \planck-like\  experiments with variable maximum multipoles using a Fisher analysis. The cutoff multipoles range from 550 to 2500 in steps of $\Delta \ell = 10$ and are shown as red points.  In each case, $\ell < 30$ is discarded and $\tau = 0.07$. There is a turning point near a cutoff multipole of $\ell = 700$. Additionally we ran MCMC runs on \planck\ data with maximum multipoles of 650, 700, 800, 1150, and 2500, which are shown as black diamonds. The turning point appears in both the MCMC fits to measured data and the Fisher analysis. \label{fig:f3}}
\end{figure}
\begin{figure}[!tbp]
  \centering
  \begin{minipage}[b]{0.4\textwidth}
    \hspace*{-1cm}
    \includegraphics[width=3.464in]{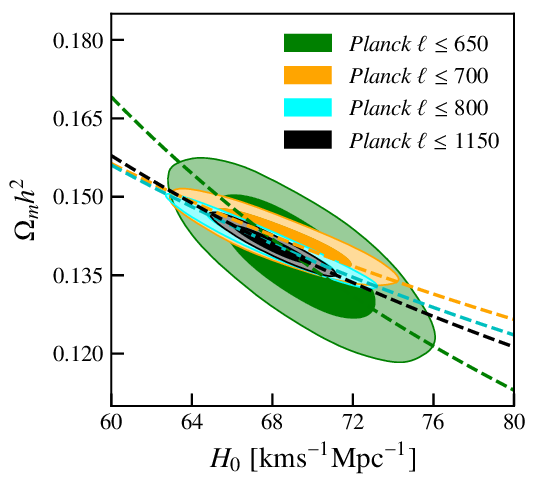}
  \end{minipage}
  \caption{ The $\Omega_mh^2-H_0$ degeneracy from MCMC runs of \planck\ data with maximum multipole cutoffs of 650, 700, 800, and 1150 are shown in green, purple, cyan, and black respectively. Additionally, dashed lines corresponding to the power law fits shown in Figure~3 are included and are colored in the same way.  The important feature of this plot is that the direction of the degeneracy changes as the Fisher calculation predicted it should with a minimum or shallowest degeneracy seen with a \planck-like\ experiment with cutoff $\ell = 700$. In each case, $\ell < 30$ was  discarded and $\tau = 0.07$. \label{fig:f4}}
\end{figure}

\begin{figure}[!tbp]
  \centering
  \begin{minipage}[b]{0.4\textwidth}
\hspace*{-0.95cm}
	\includegraphics[width=3.75in, height=3in]{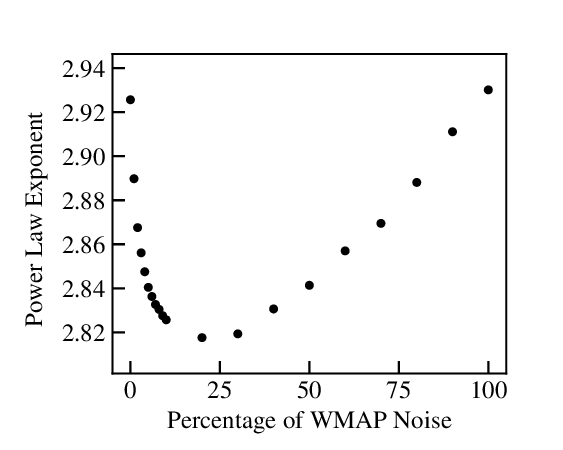}
  \end{minipage}
  \caption{ The power law exponent for a \wmap-like\ experiment varies with the level of noise in the \wmap\ multipole covariance matrix. The \wmap\ multipole covariance matrix was recomputed with different fractions of the full noise contribution and the power-law exponent was calculated in each case using the Fisher approach. Decreasing the noise contribution corresponds to increasing the maximum effective multipole measured by \wmap\ and produces a similar turning point in the exponent as observed when varying the multipole cut-off for \planck. \label{fig:f5}}
\end{figure}

\begin{figure}[!tbp]

  \begin{minipage}[b]{0.4\textwidth}
    \hspace*{-0.5cm}
    \includegraphics[width=3.75in, height=3in]{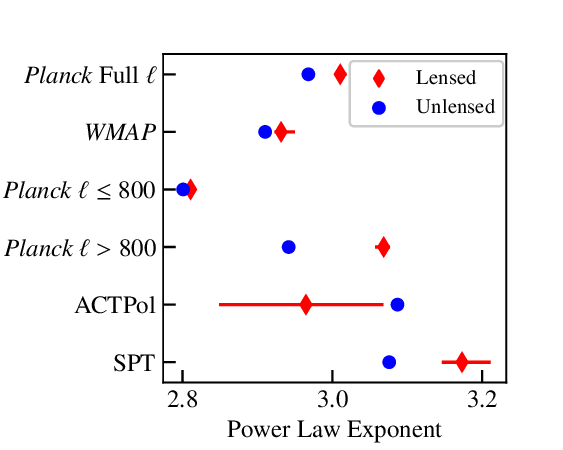}
  \end{minipage}
  \caption{We turned off lensing in \texttt{CAMB} when calculating the Fisher matrices and found the exponent $x$ that minimizes the variance of the combination $\log(\Omega_mh^x)$. For reference we include the Fisher exponents and the range corresponding to the middle 68 power law exponents from Figure 2. Note that the lensing shifts high-resolution data sets such as SPT, ACTPol, \planck\ $\ell > 800$ by an amount larger than the spread associated with possible fiducial cosmologies. We conclude that lensing plays an important role in the degeneracy direction. }
\end{figure}

The power law exponent is at a minimum at around cutoff $\ell=700$. As the maximum cutoff continues to increase, the power law exponent generally increases; however, there is a step-like behavior. The steps roughly correspond to positions of acoustic peaks with spacing $\Delta\ell\sim200$. Each step is increasingly shallower and broader, consistent with diminishing returns as further peaks are added. This basic pattern appears to be robust to assumed fiducial cosmology. Using a different fiducial model, such as the \Planck\ $\ell>800$ or  SPT best fits from Table~2, shifts the red points down or up respectively, but does not significantly alter the shape. We found the same behavior for an ideal full-sky, noiseless experiment. 

As a check of the turning point behavior, we ran \planck\ MCMC fits to the data with maximum multipoles of 650, 700, 800, 1150, and 2500. In all cases, $\tau = 0.07$ and $\ell < 30$ was discarded for comparison with the Fisher results. The agreement between the MCMC fit to data and the Fisher analysis can be seen in Figure 3. In Figure 4, we show the 1 and 2$\sigma$ confidence contours from the MCMC runs with varying cutoff. The degeneracy direction for \planck\ $\ell \leq 700$ is shallower than both \planck\ $\ell \leq 650$ and \planck\ $\ell \leq 800$, which indicates that there is a turning point. The agreement of the dashed lines representing the power law fits with the degeneracy directions of the contours indicates that the power law fit can accurately represent the data.

We also attempted to reproduce the turning point behavior with \wmap. Figure 5 shows \wmap\ Fisher calculations where the \wmap\ noise contribution is varied. While \WMAP\ measured the TT spectrum out to $\ell = 1200$, it is only cosmic variance limited for $\ell < 457$ and has a signal-to-noise ratio above unity for $\ell < 946$ \citep{bennett/etal:2013}. We modified the \wmap\ likelihood code to produce $\ell-\ell'$ multipole covariance matrices with different fractions of the full noise contribution, from 0 to $10\%$ in intervals of $1\%$, and 10 to $100\%$ in intervals of $10\%$. There is indeed a minimum in the \WMAP\ power-law exponent that occurs around $10\%$ noise with a power law exponent of 2.82. While the minimum is not exactly as low for the \planck\ forecasts, the same qualitative behavior is present. In practice, it is difficult to define an effective maximum multipole moment for each \wmap\ noise percentage; however, the \wmap\ full noise case roughly corresponds to a point along the sharp drop seen in Figure~3 at $\ell \approx 670$.

It is challenging to isolate or decouple the physical effects impacting the degeneracy between $\Omega_mh^2$ and $H_0$. One physical effect we can `turn off' is gravitational lensing, simply by calculating derivatives using unlensed $C_{\ell}^{\rm th}$ produced by \texttt{CAMB}. As shown in Figure~6, this has a significant effect on the degeneracy direction for high-$\ell$ data sets such as \Planck\ $\ell>800$ and SPT. Interestingly, the shifts in preferred power law exponent from these two data sets are larger than the spread associated with changes in the fiducial model. This implies that lensing plays an important role in the degeneracy direction for high resolution experiments in addition to peak spacing and peak heights. \Planck\ Full $\ell$ shows a smaller change, which is likely a result of the larger range of angular scales anchoring the degeneracy direction.  

There is a nonzero shift for both \planck\ $\ell \leq 800$ and \wmap\, even though the lensing effect is not significantly detected in either of these data sets. This is a consequence of not changing the fiducial \lcdm\ parameters in the Fisher analysis when lensing is turned off. We performed an MCMC fit to the \planck\ $\ell \leq 800$ data with all \lcdm\ parameters fixed, varying only the lensing amplitude with the phenomenological rescaling parameter $A_L$ \citep{calabrese/etal:2008}, and found $A_L>0$ at around $3 \sigma$. This shows that \planck\ $\ell \leq 800$ would see evidence for lensing in the case that all \lcdm\ parameters are fixed. 

ACTPol is the only data set where the Fisher-forecast power-law exponent is increased when lensing is turned off. The unlensed preferred power law exponent is within the fiducial model variance, so it is likely that this is not significant. However, this could be the result of some breakdown of the power law, which is only an approximation of the true degeneracy near the mean values of the chains. Alternatively, the Fisher matrix may break down because of non-Gaussian features in the parameter constraints.  The ACTPol polarization data are more constraining than TT data, and more precise ACTPol measurements are expected in the near future \citep{louis/etal:2017}.

\section{Conclusion}

We have quantified the $\Omega_m-H_0$ CMB degeneracy for the \planck, \wmap, ACTPol, and SPT experiments for TT power spectra within the standard \lcdm\ model. We fit the degeneracy with a power law exponent $x$ that minimizes the variance of $\log(\Omega_mh^x)$. 

We found that the CMB data sets examined preferred significantly different values of $x$, and the spread in values does not follow the preferred values of $H_0$ (Fig.~2). However, the degeneracy directions are all consistent with standard \lcdm\ model expectations, which are determined by the range of multipoles and uncertainties in the bandpower covariance matrix. 

The observed value of $x$ depends on how the multipoles are sampled and will vary with experiments. Since $\Omega_mh^x$ encompasses information about both peak heights and peak spacing, the spectrum of sensitivity to each $\ell$ affects the determination of $x$. The observed difference in preferred power law exponent between \wmap\ and \planck\ $\ell\leq 800$ is explained by a turning point in the relationship between $x$ and the maximum multipole cutoff. This turning point is the result of including the second acoustic peak, which significantly changes the degeneracies between \lcdm\ parameters. In addition to peak heights and peak spacing, gravitational lensing is a physical effect that plays a significant role in the determination of $x$ especially for high $\ell$ data (Fig~6). 

The Fisher analysis proved to be a reliable method of forecasting the degeneracy directions, and it can be used to test the consistency of current CMB experiments as well as to calculate predictions for future experiments. While we investigated only temperature data to follow Percival et al., polarization can be included in a straightforward manner.  Even though there are tensions in the preferred values of $H_0$ between data sets, such as the $2.6\sigma$ tension between SPT and \planck\ $\ell>800$ (see Figure~1 and Table~1), the physics of how the CMB is constraining the $\Omega_m-H_0$ degeneracy is consistent for all of the data sets examined. The $\Omega_m-H_0$ degeneracy directions for CMB TT data do not show evidence for the need for new physics or for the presence of significant underestimated systematics according to these tests. 
\\ 

This research was supported in part by NASA grants NNX16AF28G and NNX17AF34G. We acknowledge the use of the Legacy Archive for Microwave Background Data Analysis (LAMBDA), part of the High Energy Astrophysics Science Archive Center (HEASARC). HEASARC/LAMBDA is a service of the Astrophysics Science Division at the NASA Goddard Space Flight Center. Part of this research project was conducted using computational resources at the Maryland Advanced Research Computing Center (MARCC). The GetDist Python package was used to make Figures~1 and 3. We would like to thank Janet Weiland for her comments on a draft of this paper.

\bibliography{cosmology,planck,act,spt}

\vspace{5mm}





\end{document}